\documentclass[10pt,conference]{IEEEtran}
\usepackage{subfig}

\usepackage{hyperref}

\usepackage{graphicx}
\usepackage{xspace}
\usepackage{todonotes}
\usepackage[ruled,vlined]{algorithm2e}

\usepackage{pifont}

\newcommand{\SAP}{\textsf{\small[Company]}\xspace}
\newcommand{\VULAS}{\textsf{\small[Tool]}\xspace}
\renewcommand{\SAP}{SAP\xspace}
\renewcommand{\VULAS}{\textsf{\small Vulas}\xspace}

\newcommand{\DATASETURL}{\url{https://github.com/SAP/vulnerability-assessment-kb/tree/master/MSR2019}}

\newcommand{\toprule}{\hline}
\newcommand{\midrule}{\hline}
\newcommand{\bottomrule}{\hline}

\title{A Manually-Curated Dataset of Fixes to Vulnerabilities of Open-Source Software}

\author{
    Serena E. Ponta, Henrik Plate, Antonino Sabetta, Michele Bezzi, C\'edric Dangremont \\
    SAP Security Research \\
    Mougins, France \\
    {\tt\small \{<firstname>.<lastname>\}@sap.com}
}

\begin{document}

%
% This file was auto-generated on 2019-02-06 18:26:57.930535
%

% Number of vulnerabilities found in the NVD DEDUP
\newcommand{\ALLVULNSINNVDDEDUP}{0\xspace}

% Number of distinct vulnerabilities in the dataset 
\newcommand{\ALLVULNSCOUNT}{624\xspace}

% Number of vulnerabilities that do not have a CVE name 
\newcommand{\VULNSNOCVE}{29\xspace}

% Number of vuln with more than 10 commit (DEDUP)
\newcommand{\VULNCOMMITOVERTENDEDUP}{2\xspace}

% Number of vulnerabilities removed by the deduplication as they have the same fix commits than other vulnerabilities 
\newcommand{\VULNLOSSDEDUP}{30\xspace}

% Number of unique repositories (after removing internal, dead, svn). 
\newcommand{\REPOCOUNT}{205\xspace}

% Number of repo with 1 vuln ()
\newcommand{\REPOONEVULN}{178\xspace}

% Number of duplicate commits
\newcommand{\DUPLICATECOMMITSCOUNT}{420\xspace}

% Number of vulnerabilities found in the NVD 
\newcommand{\ALLVULNSINNVD}{0\xspace}

% Number of fix commits released in less than 100 days
\newcommand{\FIXBELOWHANDREDDAYS}{817\xspace}

% Number of repo with 1 vuln (DEDUP)
\newcommand{\REPOONEVULNDEDUP}{163\xspace}

% Number of vuln with 1 commit ()
\newcommand{\VULNONECOMMIT}{364\xspace}

% Number of vulnerabilities that do not have a CVE name DEDUP
\newcommand{\VULNSNOCVEDEDUP}{28\xspace}

% Number of distinct vulnerabilities in the dataset DEDUP
\newcommand{\ALLVULNSCOUNTDEDUP}{594\xspace}

% Number of commits 
\newcommand{\COMMITCOUNT}{1282\xspace}

% Unique vulnerabilities (after mapping vulns) DEDUP
\newcommand{\VULNCOUNTDEDUP}{594\xspace}

% Number of 'neg' commits in the dataset
\newcommand{\DATASETSIZENEG}{0\xspace}

% Number of commits in the dataset
\newcommand{\DATASETSIZEALL}{0\xspace}

% Unique vulnerabilities (after mapping vulns) 
\newcommand{\VULNCOUNT}{624\xspace}

% Number of vuln with 1 commit (DEDUP)
\newcommand{\VULNONECOMMITDEDUP}{466\xspace}

% Number of vuln with more than 10 commit ()
\newcommand{\VULNCOMMITOVERTEN}{7\xspace}

% Number of unique repositories (after removing internal, dead, svn). DEDUP
\newcommand{\REPOCOUNTDEDUP}{205\xspace}

% Number of commits DEDUP
\newcommand{\COMMITCOUNTDEDUP}{862\xspace}

% Number of 'pos' commits in the dataset (this includes duplicates!)
\newcommand{\DATASETSIZEPOS}{1282\xspace}

% Number of vulnerabilities that have a CVE but are not found in the NVD DEDUP
\newcommand{\CVENONVDDEDUP}{594\xspace}

% Number of pos commits, **de-duplicated**
\newcommand{\DATASETSIZEPOSDEDUP}{862\xspace}

% Number of vulnerabilities that have a CVE but are not found in the NVD 
\newcommand{\CVENONVD}{46\xspace}

% Number of fix commits released in less than 100 days
\newcommand{\FIXBELOWONEDAYS}{181\xspace}

\makeatletter
\begin{figure*}
\begin{minipage}{\textwidth}
\begin{center}
{\LARGE \bf \@title}\\[4mm]
{\Large [PRE-PRINT]} \\[8mm]
\@author 
\end{center}

\vspace{10mm}
\noindent\textsc{Abstract}
\input{abstract}
\vspace{15mm}
\hrule
\vspace{10mm}
\begin{center}
{\Large Citing this paper}
\end{center}

This is a pre-print of the paper that appears in the proceedings of the  16th IEEE/ACM International Conference on Mining Software Repositories (MSR), 2019.

If you wish to cite this work, please refer to it as follows:

% \begin{verbatim}
{\tt\footnotesize
@INPROCEEDINGS\{ponta2019dataset,\\
\hspace*{3mm} author=\{Serena E. Ponta and Henrik Plate and Antonino Sabetta and Michele Bezzi and C\'edric Dangremont\},\\
\hspace*{3mm} booktitle=\{Proc. of the 16th IEEE/ACM International Conference on Mining Software Repositories (MSR)\},\\
\hspace*{3mm} title=\{\@title\},\\
\hspace*{3mm} year={2019},\\
\hspace*{3mm} month={May},\\
\}
}
% \end{verbatim}
\vspace{5mm}
\hrule

\end{minipage}
\end{figure*}
\makeatother

\maketitle

\thispagestyle{empty}
\pagestyle{empty}

%%%%%%%%%%%%%%%%%%%%%%%%%%%%%%%%%%%%%%%%%%%%%%%%%%%%%%%%%%%%%%%%%%%%%%%%%%%%%%%%
\begin{abstract}

Advancing our understanding of software vulnerabilities, automating
their identification, the analysis of their impact, and ultimately their mitigation is
necessary to enable the development of software that is more secure.

While operating a vulnerability assessment tool that we developed and that is currently
used by hundreds of development units at SAP, we manually collected and curated a dataset
of vulnerabilities of open-source software and the commits fixing them. The data was obtained both
from the National Vulnerability Database (NVD) and from project-specific Web resources that we
monitor on a continuous basis. 

From that data, we extracted a dataset that maps \VULNCOUNT 
publicly disclosed vulnerabilities affecting \REPOCOUNT 
distinct open-source Java projects, used in \SAP products or internal tools,
onto the \DATASETSIZEPOS commits that fix them. Out of \VULNCOUNT vulnerabilities,
\VULNSNOCVE do not have a CVE identifier at all and \CVENONVD, which do have a CVE identifier
assigned by a numbering authority, are not available in the NVD yet.

The dataset is released under an open-source license, together with supporting scripts that allow researchers to automatically
retrieve the actual content of the commits from the corresponding repositories and to augment the attributes available for each instance. Also, these scripts allow to complement the dataset with additional instances that
are not security fixes (which is useful, for example, in machine learning applications).

Our dataset has been successfully used to train classifiers that could automatically identify security-relevant commits in code repositories. The release of this dataset and the supporting code as open-source will allow future research to be based on data
of industrial relevance; also, it represents a concrete step towards 
making the maintenance of this dataset a shared effort involving open-source communities, academia, and the industry.

%\sep{which rule do we use to decide whether to use project or repository?}
%\as{Use project if we're not referring to the source code archive specifically, otherwise use repository; if we talk about commits, it's probably going to be repository. If we're talking about industry-relevance, it's more generally the project that is relevant, not only its source code storage. Makes sense?}

\end{abstract}

%%%%%%%%%%%%%%%%%%%%%%%%%%%%%%%%%%%%%%%%%%%%%%%%%%%%%%%%%%%%%%%%%%%%%%%%%%%%%%%%
\section{Introduction}
\label{sec:intro}

% \as{Content: importance of having a vulnerability dataset; what we present, (briefly) why we made it (not how), how it differs from existing datasets~\cite{gkortzis2018msr,snyk-web,Perl:2015:VFP:2810103.2813604}}

% The adoption of open-source software (OSS) in the industry is pervasive.

While the availability of mature, high-quality open-source software (OSS)
components represents an opportunity for the software industry
to accelerate innovation and lower costs, maintaining a secure OSS supply chain and an effective vulnerability management process remains a difficult challenge.

In recent years, several approaches to the management of OSS vulnerabilities have emerged and even the market of commercial tools has matured significantly. However, as shown in~\cite{ponta2018}, an effective vulnerability management approach cannot rely purely on
metadata (which are often inaccurate, incomplete, or missing). Rather, it has to be
\emph{code-centric}, that is, it must be based on the actual analysis of vulnerabilities and their fixes at code level.

% The lack of comprehensive, consistent, and up-to-date sources of detailed (code-level) vulnerability data represents a fundamental obstacle to tackle this challenge.

% Unfortunately, the de-facto standard source of vulnerability data, the National Vulnerability Database (NVD), is known to suffer from poor coverage and varying
% data quality. Also, NVD entries (CVEs) related to OSS, fail to consistently provide a link from the advisory to the source code changes that address the vulnerability.

% Manually linking a vulnerability and the commit(s) that
% fix it requires considerable manual effort and expert knowledge, and is therefore expensive and error-prone. This problem motivated research
% on automated methods that could alleviate the cost of this activity and of vulnerability analysis in general~\cite{Perl:2015:VFP:2810103.2813604,srcclr-esecfse-2017}.

% Advancing our understanding of software vulnerabilities, and automating
% their identification, the analysis of their impact, and ultimately their mitigation is
% necessary to enable the development of software that is more secure.

While operating a vulnerability assessment tool that we developed to implement such code-centric approach, and that is currently used by hundreds of development units at SAP, we manually collected a substantial amount of code-level data about vulnerabilities of open-source software and their fixes. 

Starting from that data, and with the objective to study automated approaches to
simplify the expensive and difficult problem of manually identifying the fixes for
new vulnerability disclosures, we created a dataset, which we present here,
that was successfully used to train automated commit classifiers~\cite{sabetta2018icsme}.

Our dataset maps \VULNCOUNT publicly disclosed vulnerabilities affecting \REPOCOUNT distinct open-source Java projects used in SAP software (either products or internal tools) onto the \COMMITCOUNT commits that fix them. It was constructed and manually curated over a period of four years, monitoring the disclosure of security advisories, not only from the NVD, but also from numerous project-specific Web pages.

% Other works have been published that present datasets similar to ours, such as~\cite{gkortzis2018msr,snyk-web,Perl:2015:VFP:2810103.2813604}.

 Compared to existing works, our dataset has several distinguishing characteristics.
 \begin{itemize}
    \item Differently from~\cite{gkortzis2018msr}, who constructed their dataset by automatically extracting data from the whole NVD, we manually curated the data of  each vulnerability analyzed. While the overall amount of vulnerabilities we could cover is lower, by manually curating our dataset we could ensure the quality of each entry, detecting and addressing some of the inconsistencies that are known to affect the NVD~\cite{plate2015}, and including many commits that could not be found on the NVD. We estimate that over 70\% of the vulnerabilities in our dataset do not have any link to a source code repository in their NVD page.
    \item Because the  selection of the projects reflects the population of OSS that are actually used in real enterprise software products or internal tools, our dataset only includes projects that have practical industrial relevance. We are not aware of any other freely available dataset that has this characteristic.
    \item Differently from~\cite{Perl:2015:VFP:2810103.2813604}, our dataset contains commits that implement \emph{fixes} to vulnerabilities, whereas theirs contains commits that \emph{introduce} vulnerabilities.
    \item Snyk\footnote{\url{https://snyk.io/vuln/}}  does advertise its vulnerability databases through its website. Such database is focused on vulnerability descriptions and
    a reference to the fix is not always included.
    Also, the access to the Snyk dataset is subject to restrictions and their data are not available for free download\footnote{Snyk stopped offering dumps of their vulnerability database on GitHub as of February 2018.}. Our dataset instead is available under the Apache license and its release is part of an initiative to promote a collaborative effort involving industry, academia, and the open-source community, to maintain and extend the dataset in the long run.
\end{itemize}

In the next section, we present the process we followed to construct the dataset, whose content is described in Section~\ref{sec:dataset_description}. 
Possible applications of the dataset are illustrated in Section~\ref{sec:application}, while Section~\ref{sec:conclusion} presents some concluding remarks.

%%%%%%%%%%%%%%%%%%%%%%%%%%%%%%%%%%%%%%%%%%%%%%%%%%%%%%%%%%%%%%%%%%%%%%%%%%%%%%%%
%%%%%%%%%%%%%%%%%%%%%%%%%%%%%%%%%%%%%%%%%%%%%%%%%%%%%%%%%%%%%%%%%%%%%%%%%%%%%%%%
\section{Dataset Construction}
\label{sec:background}

The motivation for collecting code-level vulnerability data originates in the work that SAP Security Research 
has performed since 2014 on the analysis of vulnerabilities in open-source software. A key 
result of that work is an approach to the detection, assessment and mitigation of 
open-source vulnerabilities~\cite{ponta2018}. The approach is implemented as a tool
(internally called \VULAS) that has been productively used at SAP to analyze 
hundreds of Java and Python software applications and that is now publicly available as free open-source software released under the Apache version 2.0 license\footnote{\url{https://github.com/sap/vulnerability-assessment-tool}}.

The availability of a  
vulnerability knowledge base, that is comprehensive, accurate, and updated in a timely manner, is the prerequisite for such a tool to operate effectively. For this reason, we continuously monitor both the NVD and over 50 distinct project-specific websites for new vulnerability disclosures. For each such disclosure, we manually review the available information and we search for the corresponding fix-commit(s) in the code repository of the affected open-source component. The result of this ongoing activity, started in 2014, is a database of vulnerability data that has a very high coverage of the projects that are relevant to our company. From this database, we extracted \COMMITCOUNT commits, from \REPOCOUNT distinct open-source Java projects used in SAP software (either products or internal tools). These commits 
correspond to the fixes of \VULNCOUNT publicly known vulnerabilities\footnote{The dataset we describe in this paper is a snapshot of our vulnerability database taken on 21 January 2019; it is available as a comma-separated file at \DATASETURL. Some of the entries were discarded because the commits were invalid (e.g., some repositories had been dismissed or moved since we first introduced them in the dataset).}.
%\sep{"selected" sounds like we picked those we wanted. what about "retrieved" and we could also put the date 21.jan?}
%\as{I agree, but the fact is, because we discard SVN repositories, we have a dataset that is not representing a realistic population of projects SAP depends upon; saying that this is THE dataset would mean admitting we are not monitoring Tomcat, among others, which would be weird. Solution: bring back SVN, but it takes some work...}

% \sep{TO ADD here or in next section: This activity was performed on the 21 of January and resulted in the CSV file available at XXXX. }
% \sep{Do we want to say that we blacklisted something? We could anticipate the sentence below "Some of the commits turned out.."}

% around \vulascommitsapprox security-relevant commits
% corresponding to the fixes for \vulasbugsapprox vulnerabilities (e.g., CVEs).
% For our study we used a subset of 660 such commits from 152 repositories.

% The data preparation
% phase includes several steps of little technical interest, but whose
% impact on the classification results is significant: we summarize here the main ones in the hope of encouraging studies that replicate (and possibly improve) our results. 

%%%%%%%%%%%%%%%%%%%%%%%%%%%%%%%%%%%%%%%%%%%%%%%%%%%%%%%%%%%%%%%%%%%%%%%%%%%%%%%%
\section{Dataset Description}
\label{sec:dataset_description}

The dataset consists of a set of 4-tuples $$(vulnerability\_id, repository\_url, commit\_id, class)$$

where $vulnerability\_id$ is the identifier of a vulnerability that is fixed in the commit with identifier $commit\_id$ performed in the source code repository at $repository\_url$. Each entry of the dataset represents a commit that contributes to fixing a vulnerability (so-called \emph{fix commits}) and thus we assign them to the positive $class$ (\emph{pos}). In certain applications, commits of the negative class (denoted by \emph{neg}) are also needed (see Section~\ref{sec:application}). Together with the set of positive instances, we also provide supporting scripts to manipulate it and to obtain negative instances automatically. Both the data and the scripts are released under the Apache~2.0 license.

The dataset %, de-duplicated and extended as explained in the previous section, 
covers \REPOCOUNT distinct projects, and includes \COMMITCOUNT unique commits corresponding to the fixes to \VULNCOUNT vulnerabilities. Differently from other existing datasets (e.g.,~\cite{gkortzis2018msr}), our dataset includes vulnerabilities that are not available at the National Vulnerability Database, and that we obtained from project-specific advisories. In particular the dataset includes \VULNSNOCVE vulnerabilities without CVE identifier\footnote{CVE identifiers are the most-used naming convention used to enumerate vulnerabilities.}, and \CVENONVD vulnerabilities which have been given a CVE identifier by a CVE numbering authority, but are not yet published on NVD.

%\sep{As they do not do this automatically, this estimate is not meaningful anylonger, any other metrics to compare would require nasty naming mappings.}

%\sep{Do we want to also add the actual number of those with reference to git/svn? 191 }
%\as{anche no, we already mentioned this in the intro}

%Distinct CVE known to NVD: 442
%Distinct CVE names: 458 (26 do not have a CVE, 42 are not in the NVD feeds nor webapp i.e. ,reserved only just appear in the mitre and do not have any description or anything else)

The dataset can be considered to cover a representative sample of projects of practical
industrial relevance: these projects were identified based on an analysis of the data collected at SAP while operating our open-source vulnerability assessment tool (internally known as \VULAS) for a period of about four years, during which the tool was used for hundreds of thousands of security scans. Most of the open-source projects included in the dataset are hosted on \textsf{\small GitHub.com} (or a mirror of their official repository is available there).

The dataset is released as comma-separated values (CSV) file, which makes it readily usable as input to scripts that can fetch automatically additional data from each repository. We do provide example code (in the form of a Jupyter notebook and a few supporting Python scripts) that simplify the manipulation, the extension, and the analysis of the dataset.

\begin{figure}[tb]
  \centering
    \includegraphics[width=0.5\textwidth]{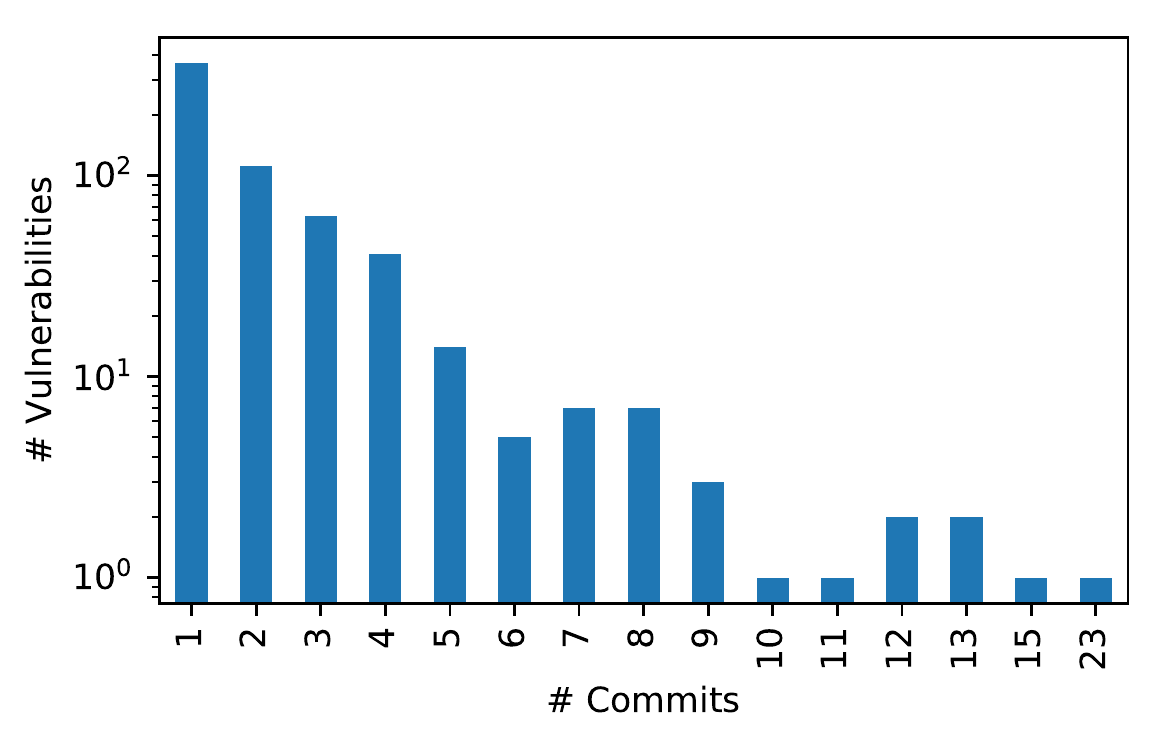}
    \vspace*{-5mm}
    \caption{Number of vulnerabilities per number of commits performed for fixing them (note: the \emph{y-axis} uses a logarithmic scale).}
    \label{fig:commitXvuln}
    \vspace*{-2mm}
\end{figure}

\begin{figure}[tb]
  \centering
    \includegraphics[width=0.5\textwidth]{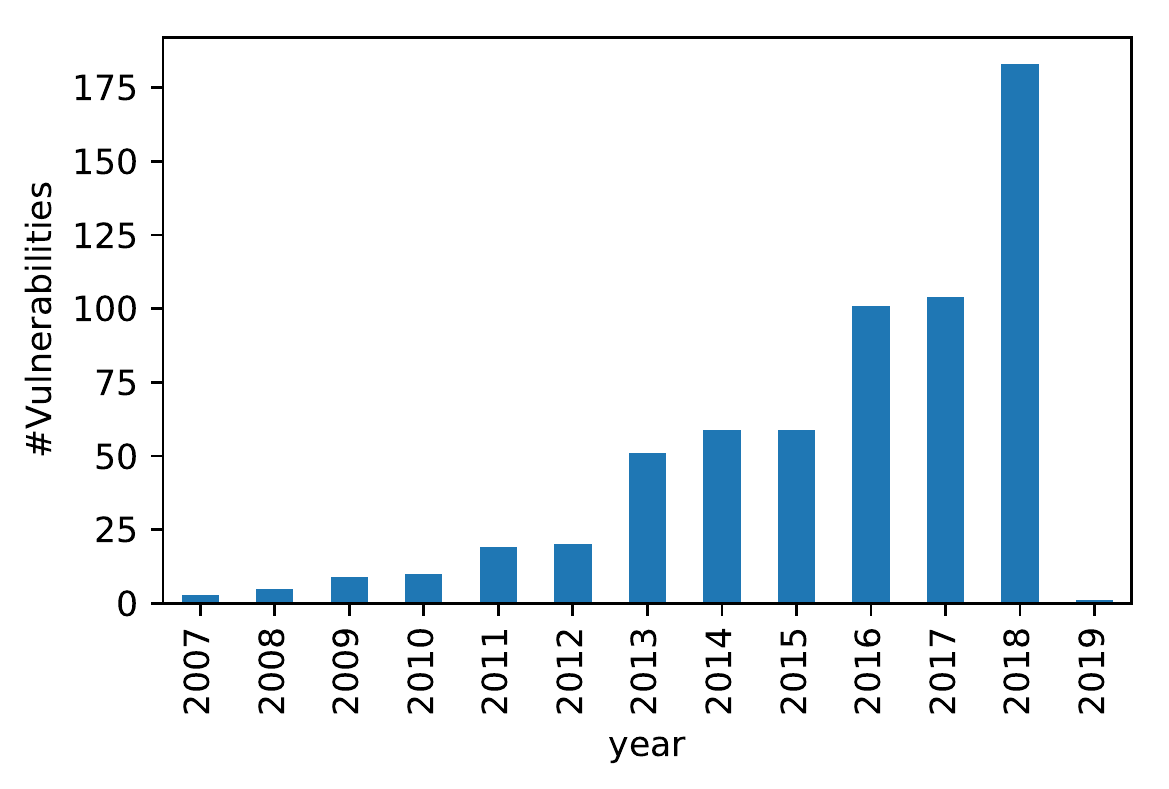}
    \vspace*{-5mm}
    \caption{Number of vulnerabilities per year.}
    \label{fig:vulnXyear}
    \vspace*{-2mm}
\end{figure}

As shown in Figure~\ref{fig:commitXvuln}, most of the vulnerabilities (\VULNONECOMMIT) are fixed in a single commit whereas in \VULNCOMMITOVERTEN cases the number of commits done to fix the vulnerability goes over ten, up to twenty-three for CVE-2015-5348. 

The existence of multiple commits per vulnerability  may indicate either that the fix is performed in steps (in which case, the entire set of commits determines the fix), or that the fix is back-ported to other versions (branches). In the latter case, the fix could be replicated as-is, or it may be adpated to make it compatible with older code.
%\sep{Should we already talk about duplicates here or is sufficient in the application section?}
The top-20 vulnerabilities by number of fix-commits is shown in Table~\ref{tab:tables}-a.

Figure~\ref{fig:vulnXyear} shows the distribution of the vulnerabilities in the dataset over the years. The bias towards recent years is due to several reasons: (1)~the number of vulnerabilities published  increases every year; (2)~often the information  available for old vulnerabilities is very limited (e.g., dead reference links, moved repositories no longer available); (3)~as the dataset is manually-curated, the effort spent to identify fix-commits needs to be allocated carefully, and we give higher priority to recent vulnerabilities.
%higher priority.  \sep{This last sentence was modified to cope with a reviewer3 comment..}

%The top-20 projects by number of fix-commits are listed in Table~\ref{tab:tables}-b
%\sep{If we want to keep this table it has to be updated, still i do not know whether we want to keep it.}

\begin{figure}[tb]
  \centering
    \includegraphics[width=0.5\textwidth]{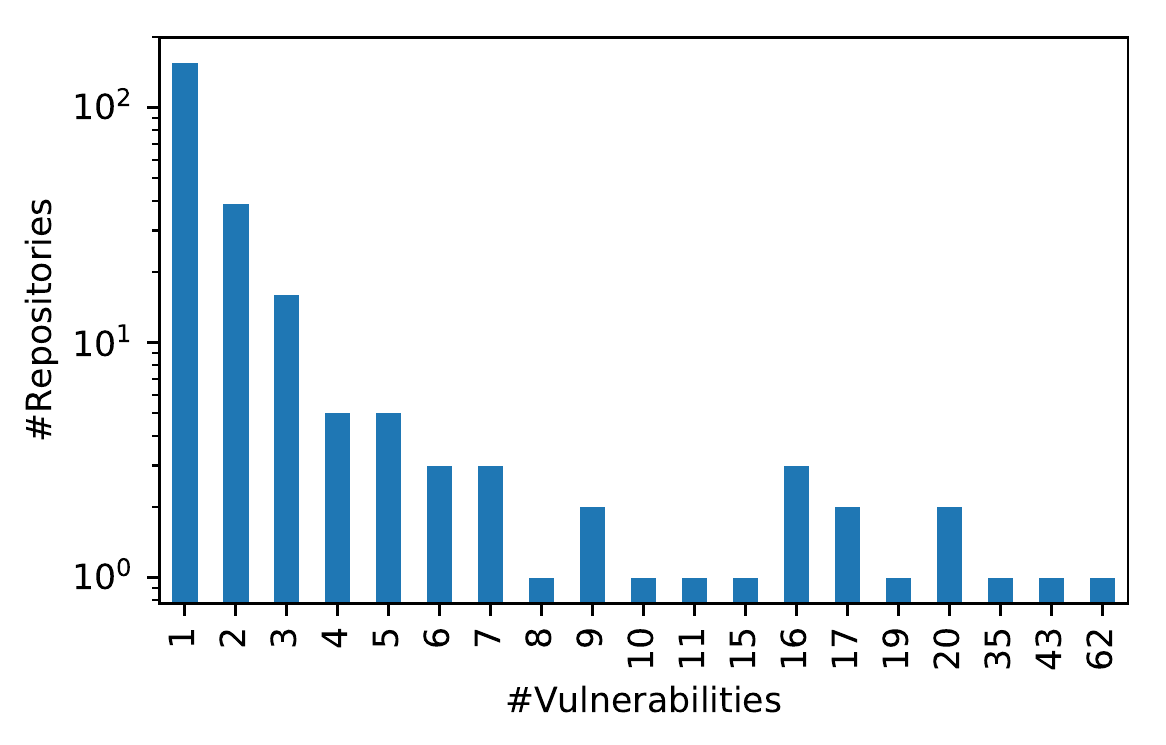}
    \vspace*{-5mm}
    \caption{Number of repositories per number of vulnerabilities (note: the \emph{y-axis} uses a logarithmic scale).}
    \label{fig:repoXvuln}
    \vspace*{-2mm}
\end{figure}

Figure~\ref{fig:repoXvuln} shows the distribution of vulnerabilities across different repositories. In particular it highlights that the majority of repositories (\REPOONEVULN over \REPOCOUNT) have only one vulnerability. The top-20 projects by number of vulnerabilities are listed in Table~\ref{tab:tables}-b

\begin{table*}[htb]
    \centering
        \subfloat[]{\begin{tabular}{lc}
\toprule
             Vulnerability Id. &  Commits \\
\midrule
       CVE-2015-5348 &       23 \\
       CVE-2012-0022 &       15 \\
       CVE-2018-8009 &       13 \\
       CVE-2016-6801 &       13 \\
       CVE-2016-8749 &       12 \\
       CVE-2018-8027 &       12 \\
       CVE-2014-0119 &       11 \\
       CVE-2012-2098 &       10 \\
       CVE-2013-1768 &        9 \\
      CVE-2017-15719 &        9 \\
       CVE-2016-5641 &        9 \\
       CVE-2016-3674 &        8 \\
       CVE-2016-5018 &        8 \\
       CVE-2014-3558 &        8 \\
       CVE-2010-0432 &        8 \\
       CVE-2016-6796 &        8 \\
      CVE-2017-12617 &        8 \\
       CVE-2017-5664 &        8 \\
      CVE-2017-15695 &        7 \\
       CVE-2015-1833 &        7 \\
       CVE-2018-1335 &        7 \\
\bottomrule
\end{tabular}
}
       % \qquad\qquad
    %   \subfloat[]{\input{data/commits_per_repo.tex}}
        \qquad
        \subfloat[]{\begin{tabular}{lc}
\toprule
                                          Repository &  Vulnerabilities \\
\midrule
                  https://github.com/apache/tomcat &             69 \\
                  https://github.com/apache/struts &             44 \\
                https://github.com/apache/tomcat70 &             35 \\
              https://github.com/jenkinsci/jenkins &             34 \\
 https://github.com/spring-projects/spring-framework &             24 \\
               https://github.com/cloudfoundry/uaa &             22 \\
                https://github.com/apache/tomcat85 &             20 \\
                https://github.com/apache/tomcat80 &             20 \\
                https://github.com/apache/activemq &             18 \\
                     https://github.com/apache/cxf &             17 \\
                https://github.com/apache/tomcat55 &             16 \\
                  https://github.com/bcgit/bc-java &             15 \\
                   https://github.com/apache/camel &             13 \\
          https://github.com/eclipse/jetty.project &             13 \\
             https://github.com/apache/lucene-solr &             11 \\
                  https://github.com/apache/hadoop &              9 \\
 https://github.com/spring-projects/spring-security &              8 \\
     https://github.com/FasterXML/jackson-databind &              8 \\
                    https://github.com/apache/tika &              7 \\
                   https://github.com/apache/ofbiz &              6 \\
                    https://github.com/apache/nifi &              6 \\
\bottomrule
\end{tabular}
}
        \qquad
        \subfloat[]{\begin{tabular}{lc}
\toprule
 Days &  \# Commits \\
\midrule
    1 &        181 \\
    2 &        166 \\
    3 &        256 \\
    4 &        259 \\
    5 &        308 \\
    6 &        282 \\
    7 &        257 \\
    8 &        136 \\
    9 &        144 \\
   10 &        172 \\
   11 &        170 \\
   12 &        129 \\
   13 &        128 \\
   14 &        512 \\
   15 &        520 \\
  15 to 50 &        592 \\
 50 to 100 &        257 \\
 100 to 500 &        290 \\
 500 to 1000 &         72 \\
1000 to 2000 &         19 \\
\emph{No tag}  &         \emph{167} \\
\bottomrule
\end{tabular}
}
    %\\[3mm]
    \vspace*{-1mm}
    \caption{Dataset description: (a)~Number of fix-commits per vulnerability (top-20); 
    %(b)~Number of fix-commits per repository (top-20); 
    (b)~Number of vulnerabilities per repository (top-20); (c)~Number of commits per days elapsed between the the fix-commit and the following tag (release).  
    }
    \vspace*{-2mm}
    \label{tab:tables}
\end{table*}

% \begin{table}
%     \centering
%     \input{data/commits_per_repo.tex}
%     \\[3mm]
%     \caption{Number of fix-commits per repository (top-20).}
%     \label{tbl:top-prj}
% \end{table}

% \begin{table}
%     \centering
%     \input{data/commits_per_vuln.tex}
%     \\[3mm]
%     \caption{Number of fix-commits per vulnerability (top-20).}
%     \label{tbl:top-prj}
% \end{table}

% The breakdown of vulnerabilities covered by year is illustrated in Figure~\ref{fig:cves-by-year}.

% \begin{figure}
%     \centering
%     % \includegraphics{}
%     \caption{Vulnerabilities by year}
%     \label{fig:cves-by-year}
% \end{figure}

% \begin{figure}
%     \centering
%     \includegraphics[width=0.43\textwidth]{figures/commits_per_vulnerability.jpg}
%     \caption{Number of unique commits per vulnerability fix.}
%     \label{fig:commits-per-vuln}
% \end{figure}

% Project coverage, statistics on patches (times, sizes, other...),
% types of vulnerabilities, other features. Some tables, plots\ldots

%%%%%%%%%%%%%%%%%%%%%%%%%%%%%%%%%%%%%%%%%%%%%%%%%%%%%%%%%%%%%%%%%%%%%%%%%%%%%%%%
\section{Applications}
\label{sec:application}

The dataset presented in this paper can be used to study vulnerabilities and their fixes in open-source software of industrial relevance (e.g.,~\cite{pashchenko2018,sabetta2018icsme}).

Using the scripts we distribute with the dataset, researchers can
easily clone all the repositories referred to in the dataset and follow our code samples to extract any commit feature available through Git.  

As an example, for each fix-commit, we automatically determine the oldest tag
from which that commit is reachable and compute the time elapsed between the
two: this can be used to study the time elapsing from the time when a
vulnerability fix is committed in a repository until a new non-vulnerable
release (containing such fix) is created (and tagged).

This can have important security consequences:
because an attacker could easily monitor issues and
commits in the repositories of security-relevant open source projects, it is
critical to keep this delay as short as possible. During that time-frame, the
fix (and thus the vulnerability that it fixes) is public, but the clients
relying on the affected open-source project cannot update to a non-vulnerable
release, because it does not exist yet.

% As a first step, we cloned all the open-source projects' repositories locally, and extracted the content (commit message, patch) of each of them.\sep{I think it can be moved to application section, right before the de-duplication.}
 
\begin{figure}[tb]
  \centering
    \includegraphics[width=0.5\textwidth]{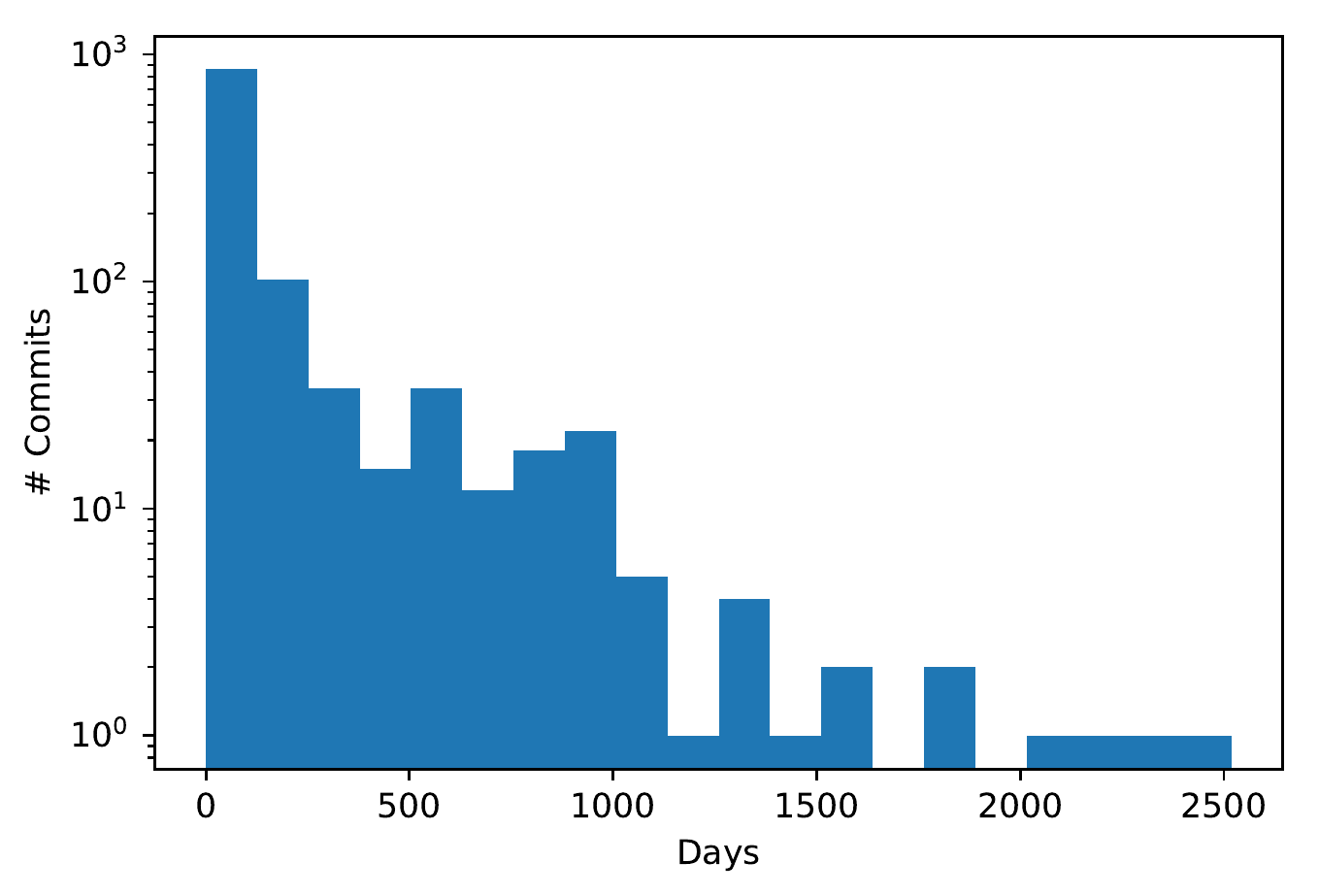}
    \vspace*{-5mm}
    \caption{Number of days from fix commit to release (note: the \emph{y-axis} uses a logarithmic scale).}
    \label{fig:commitTagDelay}
    \vspace*{-2mm}
\end{figure}

Figure~\ref{fig:commitTagDelay} shows the distribution of such delays for the commits in our dataset.
%In particular it considers the number of days passed from the commit fixing a vulnerability to the first release containing the fix, obtained by relying on the corresponding tag in the repository.
The figure shows that most fix commits (\FIXBELOWHANDREDDAYS) are released in less than 100 days (of which \FIXBELOWONEDAYS are released the same day), but delays can be much higher in a
considerable number of cases. Detailed information about how many days occur from fix-commits to their releases is available in Table~\ref{tab:tables}-c.

It is interesting to observe that as many as 167 commits are not reachable from any tag (see last line in Table~\ref{tab:tables}-c). Since most open-source projects follow the established practice of creating a tag for each release, this figure might suggest that a non-negligible number of fix-commits is available in the code repository of the project but is not yet part of any release.

%\begin{table}[htb]
%    \centering
%        \qquad\qquad
%       \subfloat[]{\input{data/delays.tex}}
%    \\[3mm]
%    \caption{Number of commits per days elapsed between %the the fix-commit and its release.  }
%    \label{tab:delays}
%\end{table}

Another example of a possible application of the dataset is presented in~\cite{sabetta2018icsme}.
Motivated by the need to automate the maintenance of the very vulnerability database from which our dataset is extracted, Sabetta and Bezzi~\cite{sabetta2018icsme} presented a novel approach to the automated classification of commits that are \emph{security-relevant} (i.e.,   that   are   likely   to   fix   a   vulnerability).
They used (an older, and smaller version of) the dataset presented here to train two independent classifiers, considering, respectively, the patch introduced by a commit (\emph{Patch Classifier}) and the \emph{log messages} (\emph{Message Classifier}), without relying on information from vulnerability advisories.

%\footnote{Of the \VULNCOUNT vulnerabilities covered by the initial data, \VULNCOUNTDEDUP are still represented by this resulting set of commits; this is due to the fact that a single commit fixes multiple vulnerabilities }.
% \sep{Should we also add the number of repositories before and after?}\sep{I think it does not become clear that the de-deuplication is done based on commit message and not same code changes. However i'm not sure we should stress it as it has non-negligible implications if we are not able to say whther the dataset has different commits with different commit msg but the same code change.}

Inspired by  the \emph{naturalness} hypothesis~\cite{allamanis2017survey,Hindle:2012:NS:2337223.2337322}, the \emph{Patch Classifier} treats source code changes as documents written in natural language (\emph{code-as-text}), and it  classifies them using established Natural Language Processing (NLP) methods. A similar approach is also used for the \emph{Message Classifier}. The results of the two classifiers, which are tuned for high precision, are then combined with a simple voting mechanism that flags a commit as security-relevant as soon as at least one of the two models does.

% The datasets introduced here (\emph{pos} and \emph{neg}) have been used for training and validation, the second field of the dataset (link) allows for the retrieval of the source code commits (\code{patch}) and corresponding log messages (\code{msg}) used by the two classifiers. 
% The combination of the two  independent classifiers, suitably tuned, is shown to yield high precision (80\%) while ensuring an acceptable recall (43\%)\footnote{Note: in an industrial scenario, the outcome goes through the assessment of a human expert; this requires that the number of false-positives be minimized,  optimizing precision at the cost of losing on recall.}.

As we pointed out in Section~\ref{sec:dataset_description}, it is common practice to commit the same change in multiple branches of the same repository,
so that the fix is made available to users of different supported versions of the component at hand.
As a consequence of this practice, our dataset contains duplicates (commits with different identifier but identical content)
which need to be removed before training a classifier.
The de-duplication results in \COMMITCOUNTDEDUP unique \emph{positive} instances.

% \noindent\textbf{Extending the dataset with instances of the `negative` class.}

Generally speaking, datasets used to train automated classifiers must include
commits corresponding both to security fixes (\emph{positive} class) and to
other changes not related to security (\emph{negative} class). To support these
applications, we provide a script that, starting from the set of positive
instances at our disposal, augments the dataset with negative instances
(non-security commits). The algorithm works as follows: for each positive
instance $p$ from repository $R$, we take $k$ random commits $n_1,\ldots,n_k$
from $R$ and, under the assumption that security-relevant commits are rare
compared to other types of commits, we treated these as ``negative'' examples.
To avoid including obvious outliers (extremely large, empty, or otherwise
invalid commits), we performed a manual review of these commits, supported by
\emph{ad-hoc} scripts and pattern matching (similarly
to~\cite{srcclr-esecfse-2017}). These patterns are used to speed-up the manual
review by searching for patterns in the commit messages that would indicate
with high probability that the commit is security-related.

The model of~\cite{sabetta2018icsme} seems to represent an improvement over a
similar state-of-the-art approach~\cite{srcclr-esecfse-2017} which relies on
log messages processed through a more complex architecture trained with a
different (and substantially larger) dataset. Unfortunately, the dataset used
in~\cite{srcclr-esecfse-2017} is not public, which makes it impossible to
conduct a reliable comparison of the two approaches.

We hope that, by sharing our dataset, we will encourage further works to
demonstrate the extent of their claimed improvements on the basis of a common
benchmark that is \emph{freely available}, \emph{machine-readable}, and that
covers open-source projects of \emph{actual industrial relevance}.

%%%%%%%%%%%%%%%%%%%%%%%%%%%%%%%%%%%%%%%%%%%%%%%%%%%%%%%%%%%%%%%%%%%%%%%%%%%%%%%%
% The rel-work material is now in the intro
% \input{related_work}
%%%%%%%%%%%%%%%%%%%%%%%%%%%%%%%%%%%%%%%%%%%%%%%%%%%%%%%%%%%%%%%%%%%%%%%%%%%%%%%%

%%%%%%%%%%%%%%%%%%%%%%%%%%%%%%%%%%%%%%%%%%%%%%%%%%%%%%%%%%%%%%%%%%%%%%%%%%%%%%%%
%\section{Limitations}
%\label{sec:limitations}
% see \url{https://antonisgkortzis.github.io/files/GMS_MSR_18.pdf}
% for sample content

%%%%%%%%%%%%%%%%%%%%%%%%%%%%%%%%%%%%%%%%%%%%%%%%%%%%%%%%%%%%%%%%%%%%%%%%%%%%%%%%
\section{Concluding remarks}
\label{sec:conclusion}

We have presented a dataset of fixes to vulnerabilities in Java OSS projects of industrial relevance, resulting from our experience with developing and operating an open-source vulnerability management solution at \SAP. 

While the data we are releasing at this time cover only Java projects, 
%(Java being the original scope of the tool for which the data were initially collected),
in the meantime the tool has evolved to support more languages and consequently we are working to extend our dataset further. 

% Second, for practical reasons, we are only releasing data about projects whose sources are stored in \textsf{\small git} repositories, although we do have data for other projects that use \textsf{\small subversion}. In many cases, a git mirror of the official svn repository could be included. Of the very few projects that we could not include in this way for this release, it is worth mentioning (because of its relevance in commercial applications) Apache Tomcat. 

% For the sake of providing a convenient way to use our dataset in practical studies,
% but also in the hope that it can be adopted as a benchmark for replication studies and in machine learning challenges, we have provided instances of both the positive the negative classes.

% Discuss the potential limitation of our approach to extend the data-set with
% negative instances. Based on the authors' direct experience in analyzing
% a large number of commits over a period of four years, we believe the probability that a randomly selected commit corresponds to a security fix to be low.

To remain useful for the validation of enterprise applications in the long term, the dataset needs to be updated on a continuous basis to  incorporate  new vulnerabilities as soon as they are disclosed. While the current dataset has been constructed at \SAP, we are working to make its future maintenance  a community effort and to encourage the very developers who  implement the security fixes for vulnerable open-source components to contribute to this dataset. To facilitate this community effort, we plan to release support tools and processes, thus complementing the release of our \emph{vulnerability assessment tool}, which was open-sourced in October 2018. 
We are convinced that the tool and its vulnerability database represent a practical contribution to ensuring the security of the open-source software supply-chain.

%%%%%%%%%%%%%%%%%%%%%%%%%%%%%%%%%%%%%%%%%%%%%%%%%%%%%%%%%%%%%%%%%%%%%%%%%%%%%%%%
% \subsection*{Acknowledgements.}  We would like to acknowledge the insightful comments of the anonymous reviewers who contributed significantly to improve this paper.
% We are also grateful to Ivan Pashchenko, Sule Kharaman, and Jamarber Bakalli for their help with the validation of our approach, and to our colleagues Henrik Plate and Serena E. Ponta for their comments on early drafts of this work.

%%%%%%%%%%%%%%%%%%%%%%%%%%%%%%%%%%%%%%%%%%%%%%%%%%%%%%%%%%%%%%%%%%%%%%%%%%%%%%%%
%
% TODO: uncomment this and pick a good lenght to balance the columns
%
% \addtolength{\textheight}{-18mm}   % This command serves to balance the column lengths

%%%%%%%%%%%%%%%%%%%%%%%%%%%%%%%%%%%%%%%%%%%%%%%%%%%%%%%%%%%%%%%%%%%%%%%%%%%%%%%%
%\section*{APPENDIX}

% \subsection*{Acknowledgments}
% \as{TODO: acknowledgements: Michele, C\'edric} 
%%%%%%%%%%%%%%%%%%%%%%%%%%%%%%%%%%%%%%%%%%%%%%%%%%%%%%%%%%%%%%%%%%%%%%%%%%%%%%%%
% \bibliographystyle{IEEEtran}
\bibliographystyle{plain}
\bibliography{main}

\end{document}